\documentclass[%
 aip,
 jmp,%
 amsmath,amssymb,
 reprint,%
]{revtex4-1}

\usepackage{graphicx}
\usepackage{dcolumn}
\usepackage{bm}
\usepackage{blindtext}
\usepackage{mathtools}
\usepackage{amsmath}

\begin{document}


\title[Variance based Computing]{
Energy-dissipation Limits in Variance-based Computing
}

\author{Sri Harsha Kondapalli}
\author{Xuan Zhang}
\author{Shantanu Chakrabartty}
\email{shantanu@wustl.edu}
\affiliation{ 
Department of Electrical and Systems Engineering, Washington University in St. Louis,\newline Missouri 63130, USA
}%

\date{\today}

\begin{abstract}
Variance-based logic (VBL) uses the fluctuations or the variance in the state of a particle or a physical quantity to represent different logic levels.
In this letter we show that compared to the traditional bi-stable logic representation the variance-based representation can theoretically achieve 
a superior performance trade-off (in terms of energy dissipation and information capacity) when operating at fundamental limits imposed by 
thermal-noise.
We show that for a bi-stable logic device the lower limit on energy dissipated per bit is 4.35KT/bit, whereas under similar operating conditions, a VBL device could achieve a lower limit of sub-KT/bit. 
These theoretical results are general enough to be applicable to different instantiations and variants of VBL ranging from digital processors based on energy-scavenging or to processors based on the emerging valleytronic devices.  
\end{abstract}

\maketitle

\begin{figure}
\centering{
\includegraphics[scale=0.37]{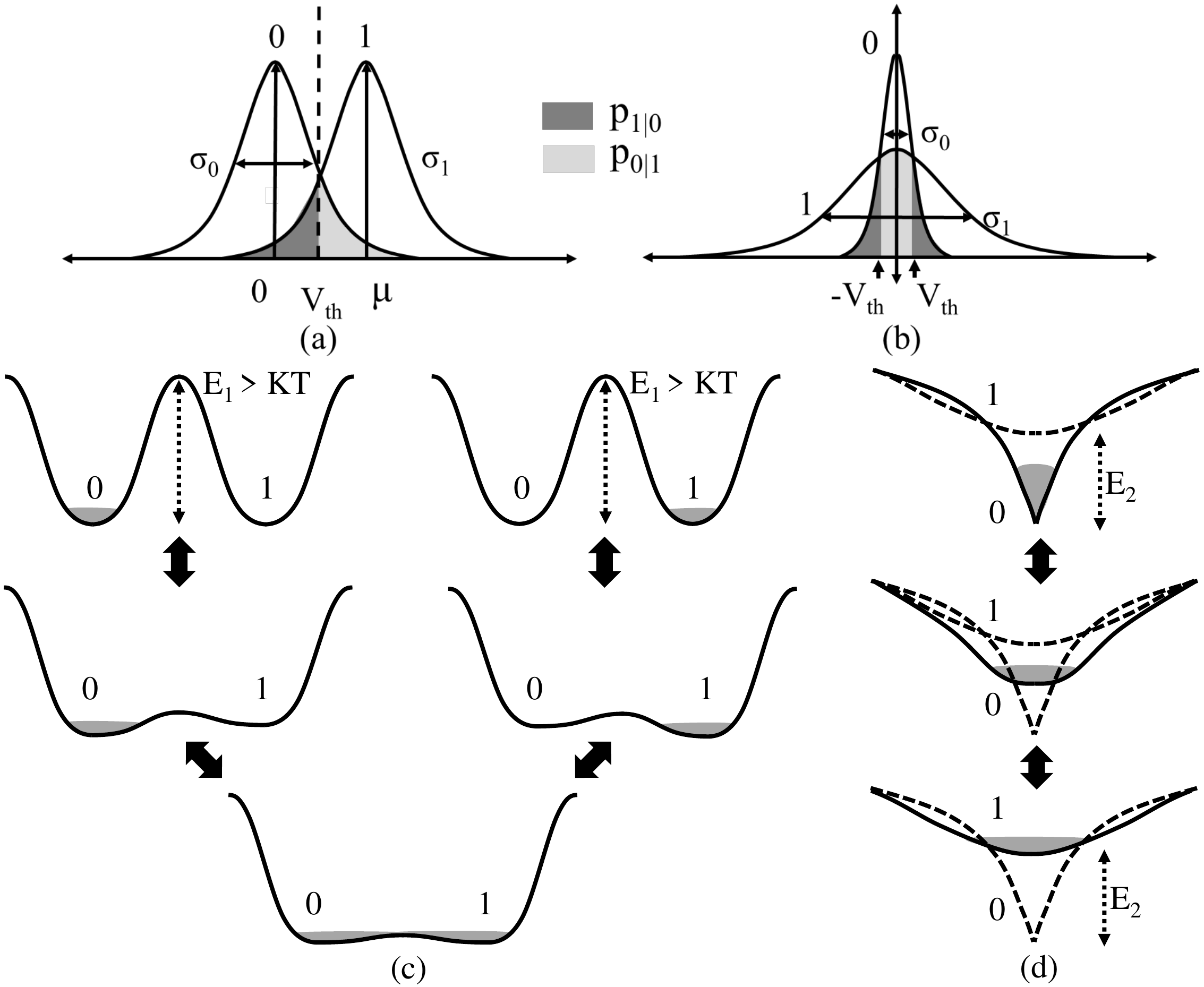}}
\caption{
Statistical representation of binary logic states '0' and '1' using (a) MBL and (b) VBL. The process of logic transition corresponding to (c) MBL and (d) VBL.
}
\label{motivation}
\end{figure}

At a fundamental level any form of digital computation involves repeated and controlled transition between different logic states. 
Traditionally, these logic states are represented and implemented using potential wells that are separated from each other
by an energy barrier. 
The potential wells are assumed to be stable configurations and any fluctuations or variance in these configurations are treated as noise. 
For example, in a standard CMOS logic the logic state is represented by the signal mean (voltage or current) and the signal variance captures the effect of thermal fluctuations or environmental interference. 
For a spintronic device\cite{SpinLogic} the logic states are represented by the state of magnetic spin of the electrons;  in a phase-change device like memristor\cite{memristor} or FeRAM\cite{FERAM} the logic states are represented by the static alignment of the molecules. 
From a statistical point of view, the two logic states (denoted by '1' and '0') are represented by the means ($0$ and $\mu$) of the two probability distributions that are separated from each other by an energy-barrier, as shown in Figure~\ref{motivation}(a). 
Note that since the physics of the two configurations are assumed to be similar, the variances of the distributions can be assumed to be equal and the probability of error can be estimated by the overlap of the distributions (shown in Fig.~\ref{motivation}(a)).
In this paper, we investigate an alternate logic representation where instead of the mean, the variances of the state configurations are used for representing logic levels $0$ and $1$.
The statistical representation of the variance-based logic (VBL)~\cite{VBLISCAS} is shown in Fig.~\ref{motivation}(b) where logic '0' is represented by a configuration with small fluctuations (or variance), and logic '1'
is represented by a configuration with large fluctuations (or variance). 
Note that in Fig.~\ref{motivation}(b), the two distributions have the same mean value which therefore does not carry any logic information. 

VBL is applicable to devices and systems where the shape of the energy levels (or equivalently the momentum of the particles) can be changed. 
One such example is a system that is powered by scavenging energy from ambient sources. In this case the asymmetry in the electrical impedance seen by the system ground and as seen by the energy transducer leads to different variances in voltage levels at the supply and at the ground potential. The difference in voltage variances could be used to implement VBL. 
Another example where VBL could be applicable are processors based on valleytronic devices~\cite{Valleytronics} where 
the curvature of the energy-levels (or equivalently the momentum of the particle trapped in the energy-level) could be changed to represent different logic levels. 
Our goal in this paper is to abstract out the physical level implementation of VBL and investigate the energy-efficiency limits of VBL as determined by thermal-noise.

In this regard, the energy-efficiency of VBL can be compared to the traditional mean-based logic (MBL) by visualizing the process of logic transition, as shown in Fig.~\ref{motivation}(c) and (d). 
For a specific implementation of MBL the logic transition can be realized by transferring electrons from one potential well to another~\cite{Landauer}$^,$\cite{bennett}, as shown in fig.~\ref{motivation}(c).  The height of the energy barrier $E_1$ which determines the reliability of a logic state is set to be at least $E_1 > KT$, where
$K$ is the Boltzmann's constant and $T$ is the temperature at which the MBL device is operated. 
During the logic transition ($0$ to $1$ for example), the energy barrier is lowered and the potential wells are reshaped in a way that the electrons move to the potential well corresponding to logic $1$. 
The energy barrier $E_1$ is then restored and held until the next transition.
Assuming irreversible computation and adiabatic transport of the electrons between the potential wells, the energy dissipated per logic transition or a bit ($E_{MBL}$) for MBL, can be estimated to be twice the height of energy barrier ($E_{MBL}=2 \times E_1$).
On the other hand, in a VBL device (as depicted in Fig.~\ref{motivation}(d)) the electrons are either constrained in a narrow potential well (a low variance state '0') or the electrons are relatively free to move around in a broader potential well (a high variance state '1'). 
Transition between the logic states in VBL involves changing the shape of the potential well and hence involves adding or subtracting a fixed amount of energy $E_2$ from the system, as shown in Fig.~\ref{motivation} (d).
Irrespective of the operating conditions for a irreversible computation the amount of energy required for each transition in case of variance based logic is approximately $E_2$ and could be significantly lower than $E_1$. 
In this paper we explore limits of the energy dissipated ($E_2$) in VBL and compare the results with equivalent bounds for MBL, similar to the bounds that have been reported in ref. (8).

\emph{Estimation of energy-dissipation per bit}: 
Following an approach similar to what was presented in ref. (8), we first estimate the information capacity for MBL and VBL by estimating the average probability of error $p_{avg}$ that is incurred in measuring the two logic levels. This can be estimated as
\begin{equation}
p_{avg} = p_0 p_{1|0} + p_1 p_{0|1}
\label{proberror}
\end{equation}
where $p_0,p_1$ are apriori probability for logic state to be '0' or '1', and $p_{1|0},p_{0|1}$ are conditional probability that captures incorrect measurement of the logic state. In an MBL representation as shown in Fig.~\ref{motivation}(a), a threshold $V_{th}$ could be used to distinguish between the logic levels in which case  
$p_{1|0},p_{0|1}$ is given by the overlap between the distributions.
Assuming equal apriori probability $p_0,p_1 = 0.5$ and the conditional distributions to be Gaussian with respective means $0$ and $\mu$ and variances $\sigma_0^2$ and $\sigma_1^2$, the average probability of error\cite{ProbabilisticCMOSpalem} can be estimated as 
\begin{equation}
\label{proberrormean}
p_{avg,MBL}=\frac{1}{4} [erfc(\frac{\mu-V_{th}}{\sqrt{2}\sigma_1})+erfc(\frac{V_{th}}{\sqrt{2}\sigma_0})]
\end{equation}
where,
\begin{equation}
\label{erfc}
erfc(x)=\frac{2}{\sqrt{\pi}}\int_{x}^{\infty} e^{-t^2} dt.
\end{equation}

In case of VBL, the variances $\sigma_0^2$ and $\sigma_1^2$ corresponding to the two logic states could be measured by comparing the magnitude of the signal with respect to a threshold $\pm V_{th}$. 
The probability of error ($p_{err,VBL}$) is determined by the shaded region as shown in Fig.~\ref{motivation} (b). 
Following equation~\ref{proberror} and assuming equal apriori probabilities,
the average probability of error $p_{err,VBL}$ can be estimated as 
\begin{equation}
\label{proberrorvar}
p_{avg,VBL}=\frac{1}{2} [1-erfc(\frac{V_{th}}{\sqrt{2} \sigma_1})+erfc(\frac{V_{th}}{\sqrt{2} \sigma_0})].
\end{equation}

The information transfer rate can be estimated by applying Shannon's capacity equation to an binary asymmetric
channel with error probabilities $p_{0|1}$ and $p_{1|0}$ and is given by 
\begin{align}
C(p_{0|1},p_{1|0})=f_c [ 1 + p_1 \{ & p_{0|1} \ln(p_{0|1}) + p_{1|1} \ln(p_{1|1}) \} + \nonumber \\ p_0 \{ & p_{1|0} \ln(p_{1|0}) + p_{0|0} \ln (p_{0|0}) \} ]
\label{asymcapacity}
\end{align}
where $f_c$ is the rate (or equivalently the speed) at which the logic state is measured. 

The next step towards determining the energy efficiency of MBL and VBL is to estimate the energy dissipated
during the process of logic transition. Similar to the approach presented in~\cite{thermalcomputing}  we will realize both the
logic by measuring an equivalent signal (mean or the variance) on an equivalent capacitance $C_{meas}$. For an
MBL, the energy is dissipated during charging and discharging the sampling capacitor 
`$C_{meas}$' to voltage $\mu$ at a rate of $f_c$ is given by
\begin{equation}
\label{powermean}
P_{MBL}=f_c \times \frac{1}{2} C_{meas} \mu^2.
\end{equation}

\begin{figure}
\centering{
\includegraphics[scale=0.44]{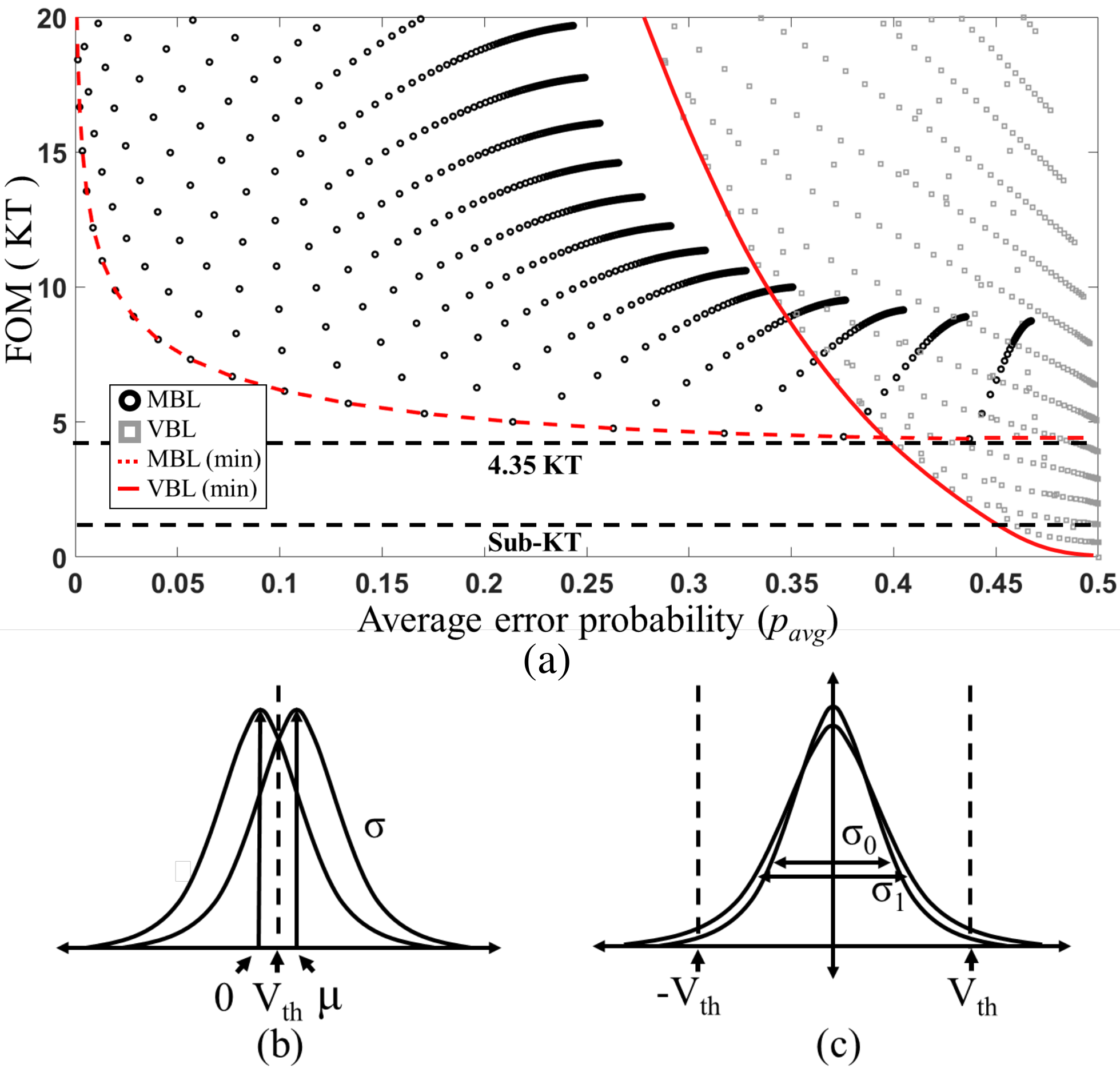}}
\caption{
(a) Numerical estimated FOM (energy dissipated per bit) corresponding to MBL and VBL for different values of $V_{th}$ and $\sigma_1$. (b)
Statistical distributions and thresholds corresponding to the (b) MBL operating at the 4.35KT per bit fundamental limit; and (c) VBL operating
at sub-KT per bit fundamental limit.}
\label{ProbvsPwr}
\end{figure}

For a VBL, the power dissipation would be given by the difference in the signal variance corresponding to the two logic states and is given by
\begin{equation}
P_{VBL}=f_c \times C_{meas}(\sigma_1^2-\sigma_0^2)
\end{equation}
The power dissipated per bit (or the figure-or-merit for comparison) is then given by 
\begin{equation}
FOM_{MBL,VBL}=\frac{P_{MBL,VBL}}{C(p_{0|1},p_{1|0})}.
\label{FOM}
\end{equation}
Note that the FOM is a function of probabilities $p_{1|0}$ and $p_{0|1}$ which in turn depend on the variances $\sigma_0^2$, $\sigma_1^2$ 
corresponding to the logic states $0$ and $1$ respectively. Since our objective is to determine the fundamental limits for MBL and VBL as
constrained by thermal noise, we will assume $\sigma_0^2 = KT/C_{meas}$. Figure~\ref{ProbvsPwr} (a) compares the FOM numerically estimated for
MBL and VBL using equations~\ref{proberrormean}-~\ref{FOM} and for different values of $V_{th}, \sigma_1^2$. The figure shows that for the FOM
for MBL is bounded from below and approaches a fundamental limit of 4.35 KT/bit. This limit is different from what was previously reported
in~\cite{thermalcomputing} and therefore in this section we provide a brief derivation of this limit.

Revisiting the approximation provided in ref. (8) it can be seen that the capacity of MBL ($C_{MBL}$), when operating near the average probability of error 
$p_{avg} = p \approx 0.5$. Assuming a binary symmetric channel  with $\sigma_1 = \sigma_0$, the Shannon capacity equation given by equation~\ref{asymcapacity} can be rewritten as 
\begin{equation}
\label{capacity}
C_{MBL}(p)=f_c[1+p\log_2p+(1-p)\log_2(1-p)].
\end{equation}

\begin{figure}
\centering{
\includegraphics[scale=0.29]{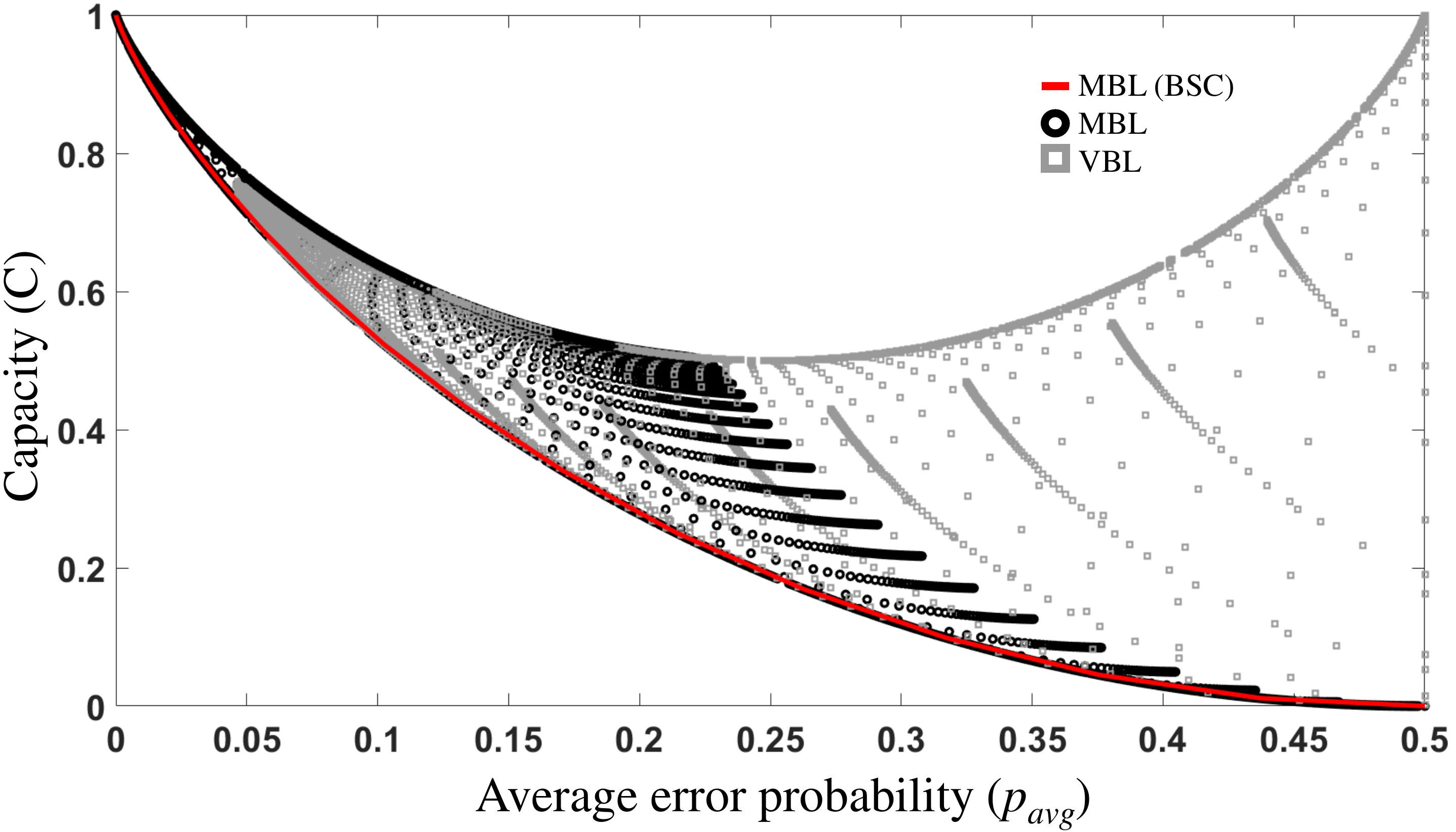}}
\caption{
Comparison of numerically estimated information capacity $C$ as a function of the average probability of error $p_{avg}$, corresponding to MBL and VBL and for different values of $V_{th}$ and $\sigma_1$.}
\label{ProbvsCap}
\end{figure}

Defining $\Delta p$ as $\Delta p = p_{avg} - 0.5$ and using a Taylor series expansion of $C_{MBL}$ around $p_{avg} =p= 0.5$,
equation~\ref{capacity} leads to,
\begin{equation}
\label{capacityapprox}
\begin{split}
C_{MBL}(\Delta p)|_{p\approx0.5} & = C(p) + \frac{C'(p)}{1!} \Delta p + \frac{C''(p)}{2!} \Delta p^2 + ...\\ 
& = \frac{2}{\ln 2} f_c (\Delta p)^2
\end{split}
\end{equation}
Assuming that the variance of measurement $\sigma_0^2 = \sigma_1^2 = \frac{KT}{C_{meas}}$ as determined by thermal-noise and $V_{th} = \frac{\mu}{2} $, 
$\Delta p$ is given by
\begin{equation}
\label{dpmean}
\Delta p \approx \frac{g(0)\mu}{2} = \frac{\mu}{2 \sqrt{2 \pi}\sigma} = \frac{\mu}{2\sqrt{2 \pi K T/C_{meas}}}
\end{equation}
where g(.) is the Gaussian distribution function. Using Eq.~\eqref{capacityapprox}, the capacity is given by 
\begin{equation}
\label{capcaitymean}
C_{MBL}|_{p\approx0.5}=\frac{\mu^2}{(4 \pi \ln 2) \frac{KT}{C_{meas}}} f_c
\end{equation}
which leads to the fundamental FOM limit as
\begin{equation}
\label{meanenergycostmean}
FOM_{MBL|min} = \frac{P_{MBL}}{C_{MBL}|_{p\approx0.5}} \approx 4.35KT/bit.
\end{equation}
This limit has been verified using numerical simulation and the results are summarized in Fig.~\ref{ProbvsPwr}(a). 
It can be also seen in Fig.~\ref{ProbvsPwr}(a) that the FOM limit for VBL could be lower than the MBL limit and in some
cases the FOM approaches sub-KT per bit. For VBL sub-KT per bit limit is achieved when the respective variances
$\sigma_0$ and $\sigma_1$ are approximately equal (implying  $p_{avg} \approx 0.5$) and the
threshold $V_{th}$ samples only the tails of the distribution, as shown in Fig.~\ref{ProbvsPwr}(c). 
To understand why VBL
can achieve sub-KT per bit limit, in Fig.~\ref{ProbvsCap} we compare the channel capacity $C(p_{0|1},p_{1|0})$ for
MBL and VBL, numerically estimated for different values of $V_{th}$ and $\sigma_1$. Fig.~\ref{ProbvsCap} shows that
while for MBL the information capacity approaches zero when the $p_{avg} \approx 0.5$, this is not the case for
some instances of VBL when $V_{th}$ is located around the tails of the distribution.  

\emph{Comparison of Signal-to-Noise Ratio for MBL and VBL}: 
In~\cite{errorcorrecting} it was proposed that one of methods to approach the fundamental limit of energy-dissipation for MBL was to use error-correcting
codes to compensate for high $p_{avg}$. A more practical approach would be to first boost the signal-to-noise ratio (SNR) of the measurement through
repeated sampling and statistical averaging.  
Given $N$ independent and identically distributed (iid) random samples $x_1$, $x_2$, ... $x_N$ from a distribution with mean $\mu$ and variance $\sigma^2$, the sample mean ($\hat{x}$) is defined as $\hat{x}$ = $\frac{\Sigma_{i=1}^{N} x_i}{N}$ and sample variance is given by $\hat{\sigma}^2$ = $\frac{\Sigma_{i=1}^{N} (x_i-\hat{x})^2}{N-1}$. 
The signal-to-noise ratio (SNR) for the measurement is given by
\begin{equation}
SNR=\frac{E[\hat{x}]^2}{E[\hat{\sigma}^2]}
\end{equation}
In case of MBL, it is given by
\begin{equation}
	\label{SNRmbl}
	SNR_{MBL}=\frac{N \mu^2}{\sigma^2}
\end{equation}
\begin{figure}
\centering{
\includegraphics[scale=0.63]{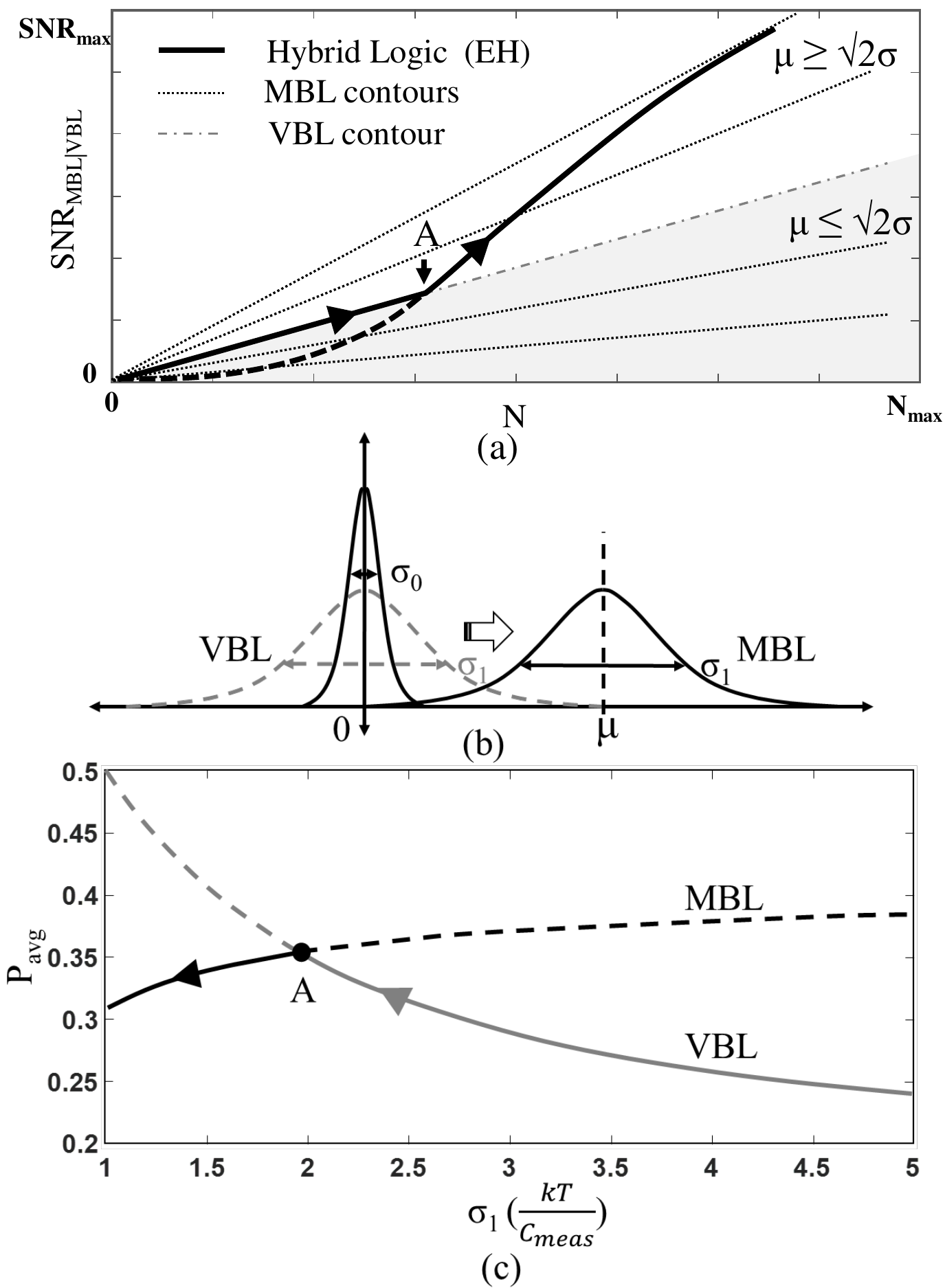}}
\caption{
Rationale for a hybrid logic that combines MBL and VBL - (a) plot showing regions where VBL (or MBL) yields a higher SNR compared to MBL (or VBL) with
'A' being the transition point between the two logic topologies; (b) Transition between VBL to MBL in an energy-scavenging system; and (c) plot showing
the variation in the average probability of error $p_{avg}$ for different values of $\sigma_1$ with 'A' being the transition point to switch between MBL and VBL. }
\label{comparison}
\end{figure}
Even if the samples are drawn from any given probability distribution the definition of $SNR_{mean}$ holds. 
Where as the variance of the sample variance becomes a function of fourth order moment and is estimated to be ~\cite{VarsamVar}
\begin{equation}
	E[(\hat{\sigma}^2 - \sigma^2)^2]=\sigma^4[\frac{2}{(N-1)}+\frac{\kappa}{N}]
\end{equation}
were $\kappa$ is the kurtosis of the probability distribution.
A generalized expression for $SNR_{var}$ is given as 
\begin{equation}
\label{SNRvbl}
	SNR_{VBL}=\frac{1}{\frac{2}{(N-1)}+\frac{\kappa}{N}}
\end{equation}

It can be seen that $SNR_{MBL}$, shown in equation.~\ref{SNRmbl}, increases with increase in $\mu$ and $N$ and with the decrease in variance ($\sigma^2$).
On the other hand $SNR_{VBL}$, shown in equation.~\ref{SNRvbl}, is independent of parameter $\sigma$ and only increases with $N$. In Fig.~\ref{comparison}(a)
we show the regions where $SNR_{VBL} \geq SNR_{MBL}$ and $SNR_{VBL} \leq SNR_{MBL}$. 

\emph{Hybrid Logic for energy scavenging processors}:
The result shown in Fig.~\ref{comparison}(a) indicates that VBL and MBL techniques could be combined to form a hybrid logic topology where VBL is used when $\mu \leq \sqrt{2}\sigma$, and MBL is used when $\mu \geq \sqrt{2}\sigma$. 
The scenario occurs in energy scavenging processors~\cite{RFlogic} where the ambient energy (for example radio-frequency signals or vibrations) could serve as source of high-variance. 
In a traditional approach, the source of energy is harvested and rectified
to create a stable voltage level $\mu$ which could then be used to implement MBL based processing. 
This process is illustrated in Fig.~\ref{comparison}(b) where during the startup phase $\mu \leq \sqrt{2}\sigma$ is satisfied and therefore VBL would be more attractive. 
As more energy is harvested and rectified, $\mu \geq \sqrt{2}\sigma$ it is more attractive to use MBL for computing.
The proposed hybrid logic should also provide improvements in reliability as the variance of the logic level $\sigma_1$ reduces in addition to the
increase in $\mu$. Fig.~\ref{comparison}(c) shows the estimated $p_{avg}$ corresponding to MBL and VBL when $\sigma_1$ is varied. The comparison
shows an optimal transition point labeled as 'A' where switching the logic style from VBL to MBL yields a better reliability in terms of $p_{avg}$.
Thus, the hybrid logic could be used to reduce system latency~\cite{VBLISCAS} and could provide significant advantages
compared to other energy harvesting logic topologies~\cite{RFlogic}$^,$\cite{AClogic}.

\end{document}